\documentclass[aps,prl,twocolumn,superscriptaddress,groupedaddress,nofootinbib]{revtex4}
\usepackage{graphicx}
\usepackage[english]{babel}
\usepackage{bm}       
\usepackage{amssymb}  
\usepackage{amsmath}
\usepackage{mathrsfs}
\usepackage{hyperref}
\usepackage{xcolor}

\begin{document}

%%%%%%%%%
\title{Phase transitions as a manifestation of spontaneous unitarity violation}

\author{ Jasper van Wezel}
\affiliation{Institute for Theoretical Physics Amsterdam, University of Amsterdam, Science Park 904, 1098 XH Amsterdam, The Netherlands}

%%%%%%%%%
\begin{abstract}
Spontaneous symmetry breaking is well understood under equilibrium conditions as a consequence of the singularity of the thermodynamic limit. How a single global orientation of the order parameter dynamically emerges from an initially symmetric state during a phase transition, however, is not captured by this paradigm. Here, we present a series of symmetry arguments suggesting that singling out a global choice for the ordered state is in fact forbidden under unitary time evolution, even in the presence of an environment and infinitesimal symmetry breaking perturbrations. We thus argue that the observation of phase transitions in our everyday world presents a manifestation of the unitarity of quantum dynamics itself being spontaneously broken. We argue that this agrees with the observation that Schr\"odinger's time dependent equation is rendered unstable for macroscopic objects owing to the same singular thermodynamic limit that affects equilibrium configurations.
\end{abstract}

\maketitle

%%%%%%%%%
\section{Dedication}
This article is dedicated to Sir Michael Berry, in celebration of his 80th birthday. It is also very much inspired by Sir Michael. The main physical concept used is the singularity of the thermodynamic limit, which I first learned to appreciate during my MSc research in Leiden, when I read Sir Michael’s excellent introduction to singular limits in Physics Today~\cite{Berry2002}. The elegantly presented ideas in that feature, and especially its parable of the apple and the maggot, inspired much of my own research since then. The description below, of the unavoidable consequences of spontaneous unitarity violations in the thermodynamic limit, is a direct application of the idea that quantum dynamics itself may be subject to a singular limit~\cite{vanWezel2008}. It results in a particular instance of the quantum to classical crossover — a topic of research that has been much influenced by Sir Michael~\cite{Berry1977,Berry1989,Berry1989scars,Berry2001}. The aspect of this crossover highlighted below first became clear to me after giving a colloquium at the University of Bristol about a related subject~\cite{vanWezel2012}. During my talk, Sir Michael asked a series of questions comparing the quantum dynamics I was describing to its classical counterpart. We did not fully agree on the differences between the two, and during dinner afterwards I continued discussing with Sir Michael. In true fashion, the matter was eloquently resolved when we agreed that in quantum mechanics, an ensemble of measurements is not necessarily the same as measuring an ensemble. The article below follows through on this argument, and presents one of its implications in the context of phase transitions and the singular thermodynamic limit.

%%%%%%%%%
\section{Introduction}
The Landau paradigm describing phase transitions in terms of the spontaneous breakdown of symmetry is one of the cornerstones of modern condensed matter physics~\cite{Landau1937,LandauBook}. The ordered, symmetry-broken state emerges from a disordered, symmetric state upon going through a phase transition, either as a function of temperature or by tuning an external control parameter. The establishment of an order parameter and the ability of the system to end up in a state with lower symmetry than its Hamiltonian are well-understood in equilibrium~\cite{vanWezel2019}, where we compare instantaneous ground states or preferred configurations without wondering how the system manages to arrive in these states. The symmetry-broken configuration is then a linear combination of symmetric states, whose collapsing tower of energy levels provide an instability or divergent susceptibility in the singular limit of infinite system size~\cite{vanWezel2019,Anderson1972,Kaiser1989,Kaplan1990,Bernu1992,Bernu1994}. 

The dynamics of phase transitions, describing the time evolution from one phase of matter into another, has likewise been studied extensively in both classical and quantum settings~\cite{Onuki2002,Sachdev2011}. This led, for example, to the modern understanding of criticality, nucleation, phase separation, defect formation, and much more. Nevertheless, we argue below that although the evolution of \emph{internal} degrees of freedom during phase transitions is well-studied, the dynamics of establishing a symmetry breaking orientation for the \emph{global} order parameter in quantum systems is not captured by these theories. In fact, we suggest that it cannot be described with any unitary time evolution governed by Schr\"odinger’s equation.  The presence of phase transitions in our everyday world is therefore a testament to the potential of the singular thermodynamic limit to generate time evolution with a lower symmetry than that dictated by the Hamiltonian~\cite{vanWezel2008}. In other words, it gives rise to an instability or divergent susceptibility in Schr\"odinger \emph{dynamics}, which results in the simultaneous spontaneous breakdown of both the symmetry of the disordered phase and the unitarity of time evolution.

Below, we first briefly review the workings of spontaneous symmetry breaking in equilibrium and emphasize the role played by the singular limits of infinite system size and vanishing perturbations. We then give a series of symmetry arguments for why the time dependent dynamics of the order parameter during a phase transition cannot be described with unitary time evolution, even in the presence of a decohering environment. Finally, we give a brief recount of a recent proposal for the emergence of spontaneous unitarity violations in the thermodynamic limit, and argue that it provides a necessary ingredient for any full description of the dynamics of phase transitions.

%%%%%%%%%
\section{Spontaneous Symmetry Breaking}
\begin{figure}[t]
\includegraphics[width=0.8\columnwidth]{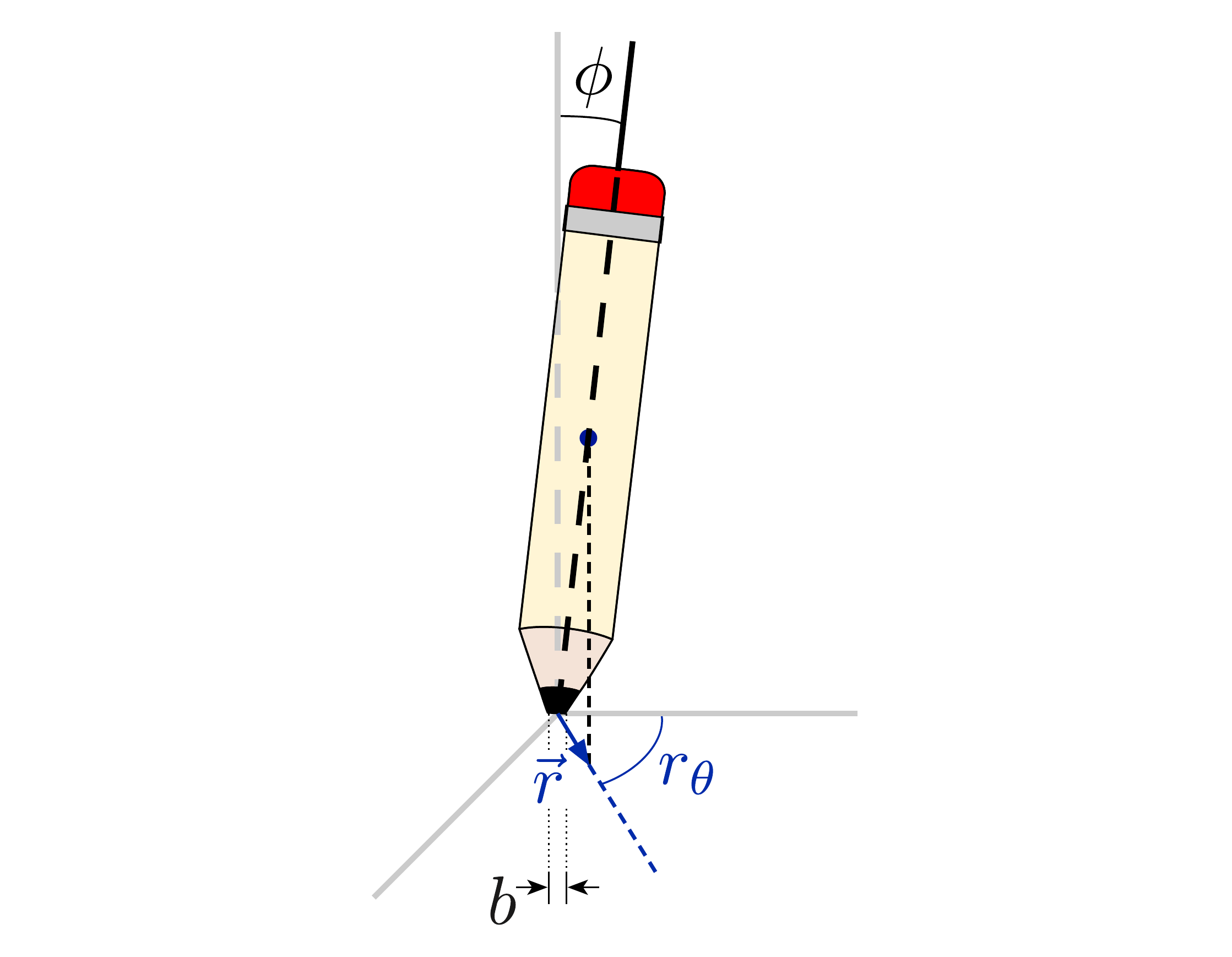}
\caption{\label{fig1} Balancing a pencil on its tip yields an instability in which the rotational symmetry can be spontaneously broken. The projection $\vec{r}$ of the center of mass on the horizontal plane acts as an order parameter. It obtains a nonzero value if the base radius $b$ is decreased to zero in the presence of any infinitesimal perturbation $\phi_0$, but is strictly zero even with only infinitesimal base area for a perfectly balanced pencil. That these limits do not commute, signals the inifinte suscpetibility of the balanced pencil towards symmetry breaking perturbations.}
\end{figure}

Recall that in classical physics, a Hamiltonian with a given symmetry may have multiple symmetry-breaking ground states that are related by the symmetry operation~\cite{vanWezel2019}. In that case, a symmetric configuration may exist as a metastable state. A classic example is that of a sharp, perfectly cylindrical pencil balancing on its tip (see figure~\ref{fig1}). The balanced state is invariant under rotations around the normal to the table, while all states of the pencil after falling flat on the table are possible ground states and these are all related by rotations around the normal. 

For the symmetric metastable state to be reduced to one of the possible ground states, it needs to break the rotational symmetry and pick a direction to fall towards. We can describe the breakdown of symmetry by considering the vector connecting the tip of the pencil to its centre of mass. Taking the projection $\vec{r}$ of that vector onto the table and writing it in polar coordinates, the angular coordinate $r_\theta$ indicates the direction into which the pencil falls and the symmetry is broken, while the the magnitude $r=|\vec{r}|$ varies between zero and a maximum value $r_{\text{max}}$ and indicates the extent to which the symmetry is broken.
%
%%%%%%%%%%%%%%%%%%%%%%%%%%
%%%%%%%%%%%%%%%%%%%%%%%%%%

The balanced state is metastable if the pencil is not infinitely sharp, as indicated by the non-zero base radius $b$ in figure~\ref{fig1}. As we consider ever sharper pencils, the balanced state becomes ever more susceptible to symmetry breaking perturbations, until finally an infinitely sharp pencil breaks the symmetry in response to even an infinitesimally small perturbation. Indicating the perturbation $\phi_0$ as a deviation of the initial state away from being perfectly balanced (as shown in figure~\ref{fig1}), the diverging susceptibility of the sharp pencil is signalled mathematically by the non-commuting nature of the two involved limits:
\begin{align}
\lim_{b\to 0} \lim_{\phi_0 \to 0} r / r_{\text{max}} &= 0 \notag \\
\lim_{\phi_0 \to 0} \lim_{b\to 0} r / r_{\text{max}} &= 1
\label{limits}
\end{align}
The fact that these limits do not commute indicates that the pencil being infinitely sharp is a \emph{singular limit}~\cite{Berry2002}. 

The infinite sensitivity to perturbations of infinitely sharp pencils makes it impossible in practice to avoid a symmetry-breaking response, and we thus say the symmetry is broken spontaneously. Notice however that actual pencils are never perfectly sharp, nor perfectly balanced, and neither of the two limits in equation~\eqref{limits} is ever realised in practice~\cite{vanWezel2019}. What these limits indicate is that it is always possible to make a pencil sufficiently sharp for it to be impossible to balance the pencil with any practical, imperfect means.

This standard picture of symmetry breaking in classical physics explains why an equilibrium symmetry-broken state is unavoidable in any practical situation. For quantum mechanical objects, the standard picture is slightly more evolved.

\subsection{Symmetry Breaking in Quantum Mechanics}
Just as in classical physics, a quantum mechanical Hamiltonian being symmetric implies that it has a ground state manifold with states related by the symmetry operation. Unlike classical physics however, the possibility for quantum systems to gain kinetic energy from tunnelling between states generically causes the ground state of quantum systems to be unique and non-degenerate~\cite{vanWezel2019}. In that case, the ground state of a quantum system is forced to carry the same symmetries as its Hamiltonian. If the balanced pencil were a quantum system, for example, its Hamiltonian might allow it to tunnel between states lying on the table at different angles. For any finite system size, there would then be a non-degenerate, symmetric ground state composed of an equal-weight superposition of all symmetry-broken states: $|\text{gs}\rangle = \int |r_\phi = \pi/2, r_\theta\rangle dr_\theta$. In this particular case, the symmetric superposition corresponds to a state of zero angular momentum.

In fact, superpositions of classical, symmetry-breaking states occur as exact ground states in quantum descriptions of almost all macroscopic objects in our everyday world, including crystals, antiferromagnets, Bose-Einstein condensates, and superconductors~\footnote{The only exceptions are ferromagnets and ferrimagnets, which have a degenerate ground state manifold and which behave like classical systems as far as symmetry breaking is concerned.}\cite{vanWezel2019,Anderson1972,Kaiser1989,Kaplan1990,vanWezel2007,Birol2007,vanWezel2008sc}. All these systems in practice choose a particular way of breaking their symmetry -- picking an angle for the quantum pencil to fall along, a position for the crystal to localise in, a direction for the antiferromagnet's sublattice magnetisation, and a value for the global phase of Bose-Einsten condensates and superconductors. In doing so, however, these quantum systems go beyond their classical analogues and realise a stable symmetry-broken state that is \emph{not} an eigenstate of their governing Hamiltonian.

This quantum version of spontaneous symmetry breaking is made possible by the presence of low-energy collective excitations in the spectrum of the symmetric Hamiltonian~\cite{vanWezel2007}. In the example of the quantum mechanical pencil, these states are labelled by the eigenvalues $l^z$ of the operator $\hat{L}^z$ for the total (global) angular momentum, which commutes with the Hamiltonian. The excitation energies of these states are proportional to $1/N$ with $N$ the number of particles making up the pencil. In the thermodynamic limit $N\to \infty$ , this tower of states collapses onto the ground state~\cite{Anderson1972}. Precisely in that limit, it then becomes possible to compose a ground state from a superposition of angular momentum eigenstates in such a way that it breaks the rotational symmetry and points along a particular angular direction: $|r_\theta\rangle = \sum_m \psi(r_\theta,m)|l_z=m\rangle$ .

For finite-sized systems, the superposition of angular momentum states corresponds to a state with minimal angular uncertainty, and is an eigenstate of the system only upon adding an external symmetry breaking field. As the number of particles increases, an ever lower strength $B$ of the symmetry breaking field is required to stabilise the minimum-uncertainty state, until in the thermodynamic limit, a true symmetry breaking state can be realised in the presence of even an infinitesimally weak field. The singularity of this limit is once again signalled by the appearance of non-commuting limits:
\begin{align}
\lim_{N\to \infty} \lim_{B \to 0} |\langle \text{gs} | e^{i \hat{\theta}} | \text{gs} \rangle| &= 0 \notag \\
\lim_{B \to 0} \lim_{N\to \infty} |\langle \text{gs} | e^{i \hat{\theta}} | \text{gs} \rangle| &= 1 
\label{SSB_gs}
\end{align}
Here, $|\text{gs}\rangle$ is the ground state of the Hamiltonian in the presence of a symmetry breaking field of strength $B$, and $\hat{\theta}$ is the angular operator canonically conjugate to $\hat{L}^z$. The operator $e^{i \hat{\theta}}$ acts as an order parameter, since it has zero expectation value in the $l^z=0$ symmetric state and obtains its maximum modulus in the symmetry broken configuration.

The tower of low-energy symmetric collective states spanning a manifold of symmetry broken configurations in the thermodynamic limit, the scaling of their energies with system size, and the resulting infinite susceptibility to symmetry breaking perturbations described by equation~\eqref{SSB_gs} are all general features of spontaneous symmetry breaking in quantum mechanics. They have been shown to stabilise the symmetry broken configuration of crystals, antiferromagnets, superconductors, and more~\cite{vanWezel2007,Birol2007,vanWezel2008sc}. It should be noted, however, that as in the case of classical symmetry breaking, neither of the two limits in equation~\eqref{SSB_gs} is ever realised in practice. What they indicate is that it is always possible to consider a quantum object that is sufficiently large for it to be impossible to avoid a symmetry broken ground state using any practical, imperfect means. 

This concludes the standard, equilibrium understanding of spontaneous symmetry breaking in both classical and quantum settings. From this well-known equilibrium basis, we can now turn attention to how symmetry-breaking configurations can be realised dynamically, as a function of time. One of the most natural places to look is in phase transitions, whose hallmark feature is the (dynamical) emergence of a symmetry-broken state from a symmetric starting point.

%%%%%%%%%
\section{Phase Transitions}
Heating a low-temperature symmetry broken system can often cause the order to melt and yield a high-temperature symmetric configuration. Conversely, starting from a high-temperature state that is invariant under symmetry transformations, a particular symmetry breaking configuration will be spontaneously chosen as the system is cooled across a phase transition. In terms of Landau's free energy expansion, this corresponds to the transition from a high-temperature free energy with a global minimum at the point of vanishing order parameter, to the system choosing one of the symmetry-related minima with nonzero order parameter at low temperatures~\cite{Landau1937,LandauBook}. Classically, which minimum is chosen in any particular experiment depends on fluctuations encountered in the exact cooling procedure. The infinite susceptibility of the classical metastable symmetric state in equilibrium can in this case be directly translated into an unavoidable dynamical evolution into one of the symmetry broken configurations. The required symmetry breaking perturbation is infinitesimally small and cannot be avoided in any practical setup.

Quantum mechanically, however, extending the understanding of why symmetry-broken states can be stable in equilibrium to understanding how they arise dynamically, is a lot less straightforward. To begin with, just the description of how to cool a quantum system requires due consideration. Entropy is strictly conserved in any closed quantum system under unitary time evolution, and so is internal energy~\footnote{The internal energy can be changed only by externally applied classical fields, which would violate the assumption of the systsem being closed once quantised.}\cite{Messiah1999}. As long as the system remains closed, a (mixed) high temperature symmetric state can therefore not be cooled to lower temperatures, let alone to a symmetry-breaking configuration.

\subsection{Open Quantum Systems}
That enacting a phase transition (dynamically, as a function of time) requires a quantum system to be open, reflects the fact that either internal energy or information needs to be dissipated in order to reach a symmetry broken state. To allow for this, consider the object of interest interacting with a second quantum mechanical system, often called the environment or bath, so that energy and information can be exchanged between the two. If the bath is sufficiently large, the energy or information flowing into it may take such a long time to return that it may be considered lost for all practical purposes~\cite{Chuang2000}. This mechanism allows a high-temperature symmetric quantum system to evolve into an entangled state with the bath, which looks mixed --and typically thermal at the bath temperature-- after the bath states have been traced out~\cite{Leggett1983,Leggett1987}.

The final step in this descricption, that of tracing out the bath, hides the true structure of the final state~\cite{Adler2003}. The coupling between system and bath should respect the symmetry of the system up to at most vanishingly small perturbations, otherwise the symmetry is broken explicitly rather than spontaneously, and a symmetry-broken state is favoured already at high temperatures. Considering first the case of purely symmetric coupling, the unitary evolution of the combined system and bath will conserve the symmetry of the system, and the final state after an arbitrarily long time of interacting with the bath can be schematically written as:
\begin{align}
|\psi\rangle = \sum_{\theta} |\theta\rangle \otimes |B_{\theta}\rangle.
\label{entangled}
\end{align}
Here, the angular variable $\theta$ labels different symmetry-broken states of the system, and the states $|B_\theta\rangle$ indicate distinct configurations of the (infinitely large) bath affected by the system being in state $|\theta\rangle$. Notice that the full state $|\psi\rangle$ is an equal-weight, symmetric, superposition over all possible symmetry-broken states $|\theta\rangle$. It also does not contain any other system states, since we assume the bath to be sufficiently large to allow the draining of all available free energy, reducing the system to its lowest-energy states only.

The final state in equation~\eqref{entangled} is an entangled state containing macroscopically distinct system and bath configurations. Tracing out the bath degree of freedom, the reduced density matrix for only the system becomes diagonal. On its own, it could then be interpreted as a mixed state with equal probability for finding any one of the symmetry broken system configurations. One can ignore the bath and use this reduced density matrix to calculate expectation values for system observables, but these can only be obtained in any actual experiment by measuring an ensemble of states with different bath configurations~\cite{Adler2003}. In each individual experiment, cooling a symmetric system across a phase transition necessarily results in a pure final state of the form of equation~\eqref{entangled} rather than singling out one of its components.

The difference between the mixed reduced density matrix and the pure full state, including entanglement with the bath, can be made tangible using an analogy in the context of the famous Schr\"odinger cat setup~\cite{Schrodinger1935}. If one places a lion (the environment) into the box alongside the cat (the system), then breaking the vial of poisonous gas would simultaneously kill both the cat and the lion, and Schr\"odinger's classic setup will yield an entangled state of the type $|\text{live},\text{live}\rangle + |\text{dead},\text{dead}\rangle$. Tracing out the lion will show the cat to have a mixed reduced density matrix, but this is unlikely to alleviate anyone's unease about the cat and lion simultaneously being both dead and alive in the full quantum state. Similarly, entangling a symmetric superposition of system states with an environment does not solve the question of how only one of the symmetry breaking states arises dynamically in any individual experiment.

\subsection{Symmetry Breaking Perturbations}
Having found that a purely symmetric coupling between system and bath does not allow for unitary dynamics from a symetric into a symmetry broken state, the question remains whether an ordered configuration may be selected by an infinitesimal symmetry breaking field. The symmetric initial state of the system can be seen as a (mixture of) equal weight superposition(s) of states with distinct orientations for the order parameter~\cite{vanWezel2019}. The coupling between system and bath can change the relative weights of these states in two ways. Starting from an initial (mixture of) product state(s) of the form $|\psi>=(\sum_\theta \psi(\theta) |\theta\rangle)\otimes|B_{\text{init}}\rangle$ the bath can cause either the values of the weights $\psi(\theta)$ to change over time, or it can cause the superposed states $|\theta\rangle$ themselves to evolve.

Considering first the case of the symmetry broken system states being eigenstates of the system-bath interaction so that only the weights $\psi(\theta)$ evolve, the reduced density matrix for the system will eventually resemble a thermal state at the bath temperature~\cite{Leggett1987,Zurek81}. For the time evolution of weights in the combined system-bath wave function to be consistent with this, the ratio of weights corresponding to differently ordered states must evolve towards the ratio of their corresponding Boltzmann weights at the bath temperature. If only the weights evolve in time, and not the states themselves, there can thus be a significant preference for a particular order parameter orientation only if the system-bath coupling causes the energies of macroscopically distinct states to differ appreciably on the scale of the critical temperature $k_\text{B}T_{\text c}$. Such large energy differences, however, would be inconsistent with the assumption that the symmetry is broken spontaneously, by an unmeasurably small perturbation.

The alternative mechanism, of the states in the initial superposition evolving in time towards a collective symmetry broken configuration, can be ruled out by a similar argument. Each of the components $|\theta\rangle$ in the initial configuration is an ordered state by itself. Rotating the order parameter of such a component towards a different orientation requires the bath to act as a canonically conjugate field to the order parameter. As a consequence of the emergent rigidity or inertia of symmetry breaking states in the thermodynamic limit, however, the time scale over which the orientation of the order parameter can evolve diverges as the strength of the conjugate field vanishes~\cite{vanWezel2019}. Once again, insisting that the bath does not explicitly break the symmetry thus implies that the weights for distinct wave function components cannot become appreciably different, ruling out the possibility of order developing spontaneously in response to an infinitesimal preference of the bath.

A final possibility is for the bath to cause \emph{local} changes in the initial system states $|\theta\rangle$, rather than a \emph{global} evolution of the collective order parameter. In classical phase transitions, thermal fluctuations may be expected to cause the appearance of small locally ordered regions within a sample, which can subsequently grow, coalesce, and eventually cause the entire system to become ordered~\cite{Onuki2002}. Quantum mechanically, all time evolution is unitary and what we interpret as thermal fluctuations in the system originates from unitarily evolving system-bath interactions. That is, for a state to be described in terms of thermal flucations in the system, the full state of system and bath combined must be an entangled superposition of the schematic form:
\begin{align}
|\psi\rangle = \sum_{\theta_1,\theta_2,\dots,\theta_N} |\theta_1,\dots,\theta_N \rangle \otimes |B_{\theta_1,\dots,\theta_N}\rangle.
\label{entangled2}
\end{align}
Here, $|\theta_1,\dots,\theta_N \rangle$ indicates a system state broken up into $N$ locally orderd regions with possibly distinct values for their local order parameters, and $||B_{\theta_1,\dots,\theta_N}\rangle$ is the corresponding bath state. Once again, the reduced density matrix for only the system may effectively look like it contains local thermal fluctuations after tracing out the bath, but the actual state of the combined system and bath in any particular experiment is necessarily a superposition over all possible system fluctuations~\cite{Adler2003}. In other words, for every system state with a local region being ordered in a particular way, there is also a component of the system-bath wave function of equation~\eqref{entangled} that has the same region ordered in the opposite way. Neither of these components can be favoured by the time evolution, and the full state of equation~\eqref{entangled} retains its overall symmetry for all time.

We thus find that all possible ways of dynamically traversing a thermal phase transition without explicitly breaking a symmetry necessarily result in an entangled system-bath state that preserves the global symmetry of the initial system state. Only upon taking an ensemble average by tracing out the environment does the effective description of the system become the expected mixture of symmetry broken configurations. For any individual evolution, the full entangled state of system and environment must be taken into account, which always remains symmetric under unitary time evolution.

\subsection{Quantum Phase Transitions}
Although the dynamics of a thermal phase transition from a symmetric to a symmetry broken state cannot be described by unitary time evolution, one might wonder if it is possible to unitarily enact a \emph{quantum} phase transition.
We thus consider two symmetric Hamiltonians, one of which allows for equilibrium spontaneous symmetry breaking in its ground state, and one that does not~\cite{Sachdev2011}. These two Hamiltonian are connected by a control parameter $p$ in such a way that each Hamitlonian is realised for a particular parameter value (say $p=0$ and $p=1$). 

If the symmetry shared by the two Hamitlonians at $p=0$ to $p=1$ is respected also for any intermediate value of the dynamically changing control parameter $p$, the order parameter associated with the symmetry is a conserved quantity throughout the evolution, and the symmetry is never broken~\cite{Messiah1999}. Moreover, even in the presence of a non-zero symmetry breaking field, starting from a configuration with a nonzero but vanishingly small order parameter means that the initial state is a superposition of the form $|\psi\rangle = \sum_{\theta} \psi(\theta) |\theta\rangle>$ of many symmetry-related states with vanishingly small differences in weight. These differences cannot be amplified by any finite amount using an infinitesimal symmetry-breaking field, owing to the rigidity or inertia that symmetry broken states attain in the thermodynamic limit~\footnote{This same inertia prevents already-localised or symmetry broken macroscopic objects in our everyday world from spontaneously tunnelling to a different location or symmetry broken state.}\cite{vanWezel2019}.

We thus find that upon tuning the Hamiltonian of a closed quantum system across a phase transition (from $p=0$ to $p=1$), the state of the system has no way of unitarily evolving into a symmetry-broken configuration. That is, the observation of a single symmetry-broken state being dynamically realised in practice during a single, individual phase transition, is at odds with the unitarity of quantum mechanical time evolution, both for thermal and quantum phase transitions.

%%%%%%%%%
\section{Spontaneous Unitarity Violation}
In all scenarios discussed so far, it is the unitarity of quantum mechanical time evolution which prevents the dynamical emergence of a symmetry broken state, even in the presence of an infinitesimal symmetry breaking perturbation and an external bath. It has been pointed out however that the unitary symmetry of the time-dependent Schr\"odinger equation may itself be spontaneously broken~\cite{vanWezel2008,vanWezel2010,Mertens2021}. Since unitarity is a property of time evolution, its breakdown emerges in the dynamics of quantum states rather than their equilibrium configurations, but the ingredients for unitarity breaking are the same as those for equilibrium symmetry breaking: a singular thermodynamic limit and an infinitesimal symmetry-breaking field. 

Time evolution in quantum mechanics is unitary because it is generated by the Hamiltonian, which, like every quantum mechanical observable, is a Hermitian operator. In order to have non-unitary time evolution, its generator needs to be non-Hermitian:
\begin{align}
i \hbar \frac{d |\psi\rangle}{dt} = (\hat{H} - i \epsilon \hat{B}) |\psi\rangle 
\label{nonuni}
\end{align}
Here, $\epsilon\hat{B}$ is a Hermitian operator of strength $\epsilon$, so that $i \epsilon \hat{B}$ is anti-Hermitian and the time evolution is rendered non-unitary. If the perturbation $i \epsilon \hat{B}$ couples to an extensive property of the system described by $\hat{H}$, it will cause evolution on a time scale proportional to $1/(\epsilon N)$, and may thus have an instantaneous effect in the thermodynamic limit even for infinitesimal strength $\epsilon$ of the unitarity-breaking field~\cite{vanWezel2008,vanWezel2010}. 

Notice that taken as a whole, $\hat{H} - i \epsilon \hat{B}$ is reminiscent of non-Hermitian Hamiltonians used to describe the effective dynamics of open quantum systems or the time evolution of driven and dissipative classical systems~\cite{Berry1998,Hatano1997,Wang2018,Alvarez2018}. Here, we do not interpret the non-Hermitian operator to describe the effective (environment-averaged) dynamics of the system, but rather we consider $B$ to be an actual non-Hermitian perturbation to Schr\"odinger's equation which is present even for a single, isolated, closed system~\cite{vanWezel2010,Mertens2021}. Moreover, halmark effects of non-Hermitian Hamiltonians, such as the non-Hermitian skin effect~\cite{Lee2019,Bergholtz2020,Ghatak2020} are neglicible in the present analysis because the localization length of the skin modes scales inversely with $\epsilon$, which we take to be infinitesimal.

Furthermore, notice that $B$ may render the combination $\hat{H} - i \epsilon \hat{B}$ PT-symmetric~\cite{Bender1998,Mostafazadeh2002,Lara2015}. The fact that such a PT-symmetry can ensure the spectrum to be real, however, is irrelevant to the present discussion, because infinitesimal values of $\epsilon$ allow only infinitesimal deviations from real eigenvalues. We also do not introduce an alternative, pseudo-Hermitian inner product that could restore unitary time evolution~\cite{Sudarshan1961,Scholz1992}, because we are interested in the spontaneous breakdown of unitarity.

For a system $\hat{H}$ that can undergo spontaneous symmetry breaking in equilibrium, we may consider $\hat{B}$ to equal the equilibrium symmetry breaking field. In that case, the non-unitary perturbation generates transitions between states whose difference in energy (defined as $\langle \hat{H} \rangle$) is of order $1/N$, and the non-unitary dynamics is energy-conserving for all practical purposes. The lack of a conserved norm for the wave function likewise does not pose a problem, since one can define the expectation value of an observable $\hat{O}$ as $\langle \psi(t) | \hat{O} | \psi(t)\rangle / \langle \psi(t) |\psi(t) \rangle$ without affecting any of the predictions of standard quantum mechanics~\cite{Mertens2021}. The dual effect of the unitarity and symmetry breaking field $\hat{B}$ in this case, is to single out a particular symmetry broken configuration and to stabilise it by dynamically suppressing the weight of all other states in the initial state superposition. 

\begin{figure}[t]
\includegraphics[width=\columnwidth]{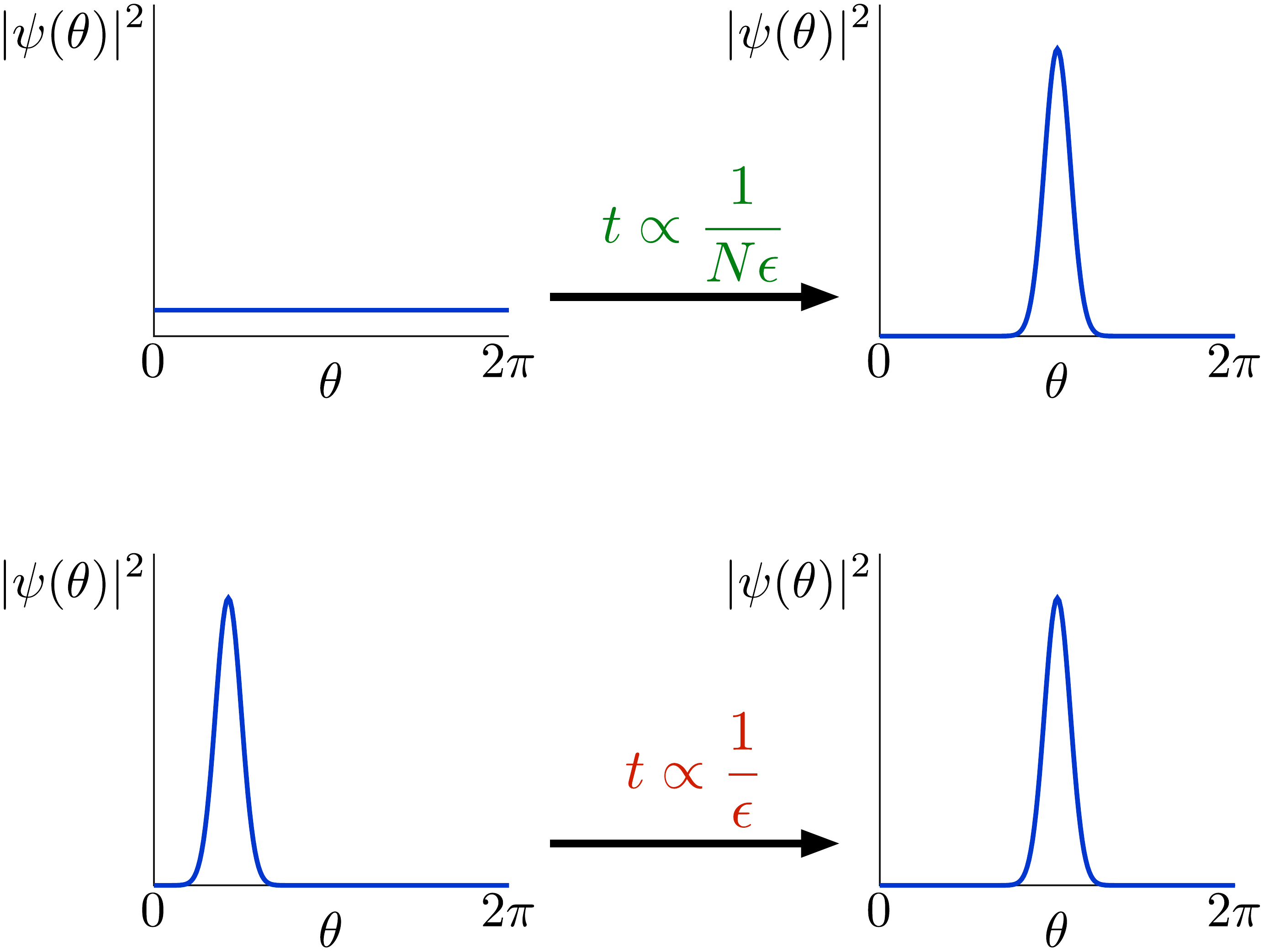}
\caption{\label{fig2} The time evolution of quantum states in the presence of a unitarity breaking perturbation. If the non-unitary perturbation has strenth $\epsilon$ and couples to the order parameter for a quantum system of size $N$, its effect on an initial state superposed over multiple directions for the order parameter (top left) will be to reduce it to a single symmetry broken state (top right) within a time scaling as $1/(N\epsilon)$. Acting on an initial state that is already ordered on the other hand (bottom left), yields a state with a different order parameter orientation only after a time scaling as $1/\epsilon$. In the thermodynamic limit of infinite system size, the former process becomes instantaneous even for infinitesimal values of $\epsilon$, while the latter process is arbitrarily slow. That the limits $N\to\infty$ and $\epsilon\to 0$ do not commute signals the inifinite susceptibility of macroscopic quantum objects towards unitarity breaking perturbations.}
\end{figure}

Precisely for the case of systems with a spontaneously broken symmetry in equilibrium, the time evolution imposed by $\hat{B}$ is singular. As shown in figure~\ref{fig2}, it reduces a general initial state to just a single symmetry broken configuration in a time proportional to $1/(\epsilon N)$, which becomes instantaneous in the thermodynamic limit even for infinitesimal field strength. That is, the limits $N\to\infty$ and $\epsilon\to 0$ fail to commute (as they did in equation~\eqref{limits}). At the same time, starting from a single symmetry broken state, $\hat{B}$ has an effect only on a time scale proportional to $1/\epsilon$, which is arbitrarily long regardless of system size. Together, these two properties show that the time evolution of systems that can break a symmetry in equilibrium is \emph{unstable}~\cite{vanWezel2008,vanWezel2010,Mertens2021}. They have an infinite sensitivity for unitarity breaking perturbations that instantaneously reduce the state to a symmetry broken configuration, which is then stable under further non-unitary perturbations. 
%
%%%%%%%%%%%%%%%%%%%%%%%%%%
%%%%%%%%%%%%%%%%%%%%%%%%%%

The infinitesimal non-unitary perturbation thus accomplishes what interactions with an environement and infinitesimal Hermitian perturbations to the Hamiltonian could not: they allow a symmetric initial state to \emph{dynamically} break its symmetry and develop a non-zero order parameter as time evolves.

%%%%%%%%%
\section{Discussion}
Just as in equilibrium symmetry breaking, the singular thermodynamic limit $N\to \infty$ does not exist in any real systems, and we can therefore expect to observe spontaneous non-unitary dynamics only if $\epsilon$ is actually nonzero. Notice that only an exceedingly small value of $\epsilon$ is required to stabilise an ordered state in macroscopic objects consisting of $\sim 10^{23}$ particles, but it must nevertheless not be identically zero. Since all observables in quantum mechanics are Hermitian, and time evolution is defined to be generated by the observable $\hat{H}$, non-unitary perturbation are forbidden in standard quantum mechanics by axiom. Equation~\eqref{nonuni} should thus be seen as a perturbative modification of Schr\"odinger's equation, going beyond quantum theory~\cite{vanWezel2010}. The underlying idea is that like every other theory in physics, quantum mechanics may be expected to have a limited domain of applicability, and the non-unitary term $i\epsilon \hat{B}$ represents the first order effect of physical laws applicable to realms where quantum theory does not apply. As an example, one possible candidate for a realm beyond the quantum is given by general relativity, whose famous incompatibility with quantum mechanics has been argued before to stem from its non-unitary nature~\cite{Penrose}.

Regardless of its precise origin, the instability of Schr\"odinger dynamics indicates that as long some component of the non-unitary corrections to quantum mechanics couple to an order parameter with vanishingly small but nonzero strength, the instantaneous emergence of symmetry broken states in macroscopic objects is unavoidable. 

During the \emph{dynamics} of an individual phase transition, an ordered, symmetry broken state arises from a disordered, symmetric state as a function of time. We argued that such a dynamical breakdown of symmetry cannot be realised under unitary time evolution as long as the system, bath, and system-bath coupling do not explicitly (i.e. with non-vanishing strength) break the symmetry. Infinitesimal preferences for any particular ordered state cannot induce a spontaneous dynamical breakdown of symmetry, because even in the thermodynamic limit the unitarity of time evolution requires all states with distinct order parameter orientations to appear as components with at most infinitesimal differences in weight in the final state wave function. The experimental observation that individual phase transitions are realised in our everyday world, are traversed as a function of time, and do result in a single symmetry-broken state each time they are encountered, directly suggests the existence of non-unitary time evolution. The theoretical observation that Schr\"odinger dynamics is intrinsically unstable in the thermodynamic limit then provides a tantalising suggestion for how non-unitary dynamics results in the emergence of a single order parameter orientation upon dynamically traversing a phase transition.

%%%%%%%%%

%%%%%%%%%

\begin{thebibliography}{42}
\expandafter\ifx\csname natexlab\endcsname\relax\def\natexlab#1{#1}\fi
\expandafter\ifx\csname bibnamefont\endcsname\relax
  \def\bibnamefont#1{#1}\fi
\expandafter\ifx\csname bibfnamefont\endcsname\relax
  \def\bibfnamefont#1{#1}\fi
\expandafter\ifx\csname citenamefont\endcsname\relax
  \def\citenamefont#1{#1}\fi
\expandafter\ifx\csname url\endcsname\relax
  \def\url#1{\texttt{#1}}\fi
\expandafter\ifx\csname urlprefix\endcsname\relax\def\urlprefix{URL }\fi
\providecommand{\bibinfo}[2]{#2}
\providecommand{\eprint}[2][]{\url{#2}}

\bibitem[{\citenamefont{Berry}(2002)}]{Berry2002}
\bibinfo{author}{\bibfnamefont{M.~V.} \bibnamefont{Berry}},
  \bibinfo{journal}{Physics Today} \textbf{\bibinfo{volume}{55}},
  \bibinfo{pages}{10} (\bibinfo{year}{2002}).

\bibitem[{\citenamefont{van Wezel}(2008)}]{vanWezel2008}
\bibinfo{author}{\bibfnamefont{J.}~\bibnamefont{van Wezel}},
  \bibinfo{journal}{Phys. Rev. B} \textbf{\bibinfo{volume}{78}},
  \bibinfo{pages}{054301} (\bibinfo{year}{2008}).

\bibitem[{\citenamefont{Berry}(1977)}]{Berry1977}
\bibinfo{author}{\bibfnamefont{M.~V.} \bibnamefont{Berry}},
  \bibinfo{journal}{Phil. Trans. R. Soc. A} \textbf{\bibinfo{volume}{287}},
  \bibinfo{pages}{237} (\bibinfo{year}{1977}).

\bibitem[{\citenamefont{Berry}(1989{\natexlab{a}})}]{Berry1989}
\bibinfo{author}{\bibfnamefont{M.~V.} \bibnamefont{Berry}},
  \bibinfo{journal}{Phys. Scripta} \textbf{\bibinfo{volume}{40}},
  \bibinfo{pages}{335} (\bibinfo{year}{1989}{\natexlab{a}}).

\bibitem[{\citenamefont{Berry}(1989{\natexlab{b}})}]{Berry1989scars}
\bibinfo{author}{\bibfnamefont{M.~V.} \bibnamefont{Berry}},
  \bibinfo{journal}{Proc. R. Soc. A} \textbf{\bibinfo{volume}{423}},
  \bibinfo{pages}{219} (\bibinfo{year}{1989}{\natexlab{b}}).

\bibitem[{\citenamefont{Berry}(2001)}]{Berry2001}
\bibinfo{author}{\bibfnamefont{M.~V.} \bibnamefont{Berry}}, in
  \emph{\bibinfo{booktitle}{Quantum mechanics: Scientfic perpectives on Divine
  Action}}, edited by \bibinfo{editor}{\bibfnamefont{R.~J.}
  \bibnamefont{Russell}},
  \bibinfo{editor}{\bibfnamefont{P.}~\bibnamefont{Clayton}},
  \bibinfo{editor}{\bibfnamefont{K.}~\bibnamefont{Wegter-McNelly}},
  \bibnamefont{and}
  \bibinfo{editor}{\bibfnamefont{J.}~\bibnamefont{Polkinghorne}}
  (\bibinfo{publisher}{Vatican Observatory – CTNS Publications},
  \bibinfo{year}{2001}), p.~\bibinfo{pages}{41}.

\bibitem[{\citenamefont{Oosterkamp and van Wezel}(2012)}]{vanWezel2012}
\bibinfo{author}{\bibfnamefont{T.~H.} \bibnamefont{Oosterkamp}}
  \bibnamefont{and} \bibinfo{author}{\bibfnamefont{J.}~\bibnamefont{van
  Wezel}}, \bibinfo{journal}{Proc. R. Soc. A} \textbf{\bibinfo{volume}{468}},
  \bibinfo{pages}{35} (\bibinfo{year}{2012}).

\bibitem[{\citenamefont{Landau}(1937)}]{Landau1937}
\bibinfo{author}{\bibfnamefont{L.~D.} \bibnamefont{Landau}},
  \bibinfo{journal}{Zh. Eksp. Teor. Fiz.} \textbf{\bibinfo{volume}{7}},
  \bibinfo{pages}{19} (\bibinfo{year}{1937}), \bibinfo{note}{translation:
  \emph{Ukr. J. Phys.} {\bf 53}, 25 (2008)}.

\bibitem[{\citenamefont{Landau and Lifshitz}(2013)}]{LandauBook}
\bibinfo{author}{\bibfnamefont{L.~D.} \bibnamefont{Landau}} \bibnamefont{and}
  \bibinfo{author}{\bibfnamefont{E.~M.} \bibnamefont{Lifshitz}},
  \emph{\bibinfo{title}{Statistical Physics}} (\bibinfo{publisher}{Elsevier},
  \bibinfo{year}{2013}), vol.~\bibinfo{volume}{5} of
  \emph{\bibinfo{series}{Course of Theoretical Physics}},
  chap.~\bibinfo{chapter}{14}, \bibinfo{edition}{3rd} ed., ISBN
  \bibinfo{isbn}{9780080570464}.

\bibitem[{\citenamefont{Beekman et~al.}(2019)\citenamefont{Beekman, Rademaker,
  and van Wezel}}]{vanWezel2019}
\bibinfo{author}{\bibfnamefont{A.~J.} \bibnamefont{Beekman}},
  \bibinfo{author}{\bibfnamefont{L.}~\bibnamefont{Rademaker}},
  \bibnamefont{and} \bibinfo{author}{\bibfnamefont{J.}~\bibnamefont{van
  Wezel}}, \bibinfo{journal}{SciPost Phys. Lect. Notes}
  \textbf{\bibinfo{volume}{11}} (\bibinfo{year}{2019}).

\bibitem[{\citenamefont{Anderson}(1972)}]{Anderson1972}
\bibinfo{author}{\bibfnamefont{P.~W.} \bibnamefont{Anderson}},
  \bibinfo{journal}{Science} \textbf{\bibinfo{volume}{177}},
  \bibinfo{pages}{393} (\bibinfo{year}{1972}).

\bibitem[{\citenamefont{Kaiser and Peschel}(1989)}]{Kaiser1989}
\bibinfo{author}{\bibfnamefont{C.}~\bibnamefont{Kaiser}} \bibnamefont{and}
  \bibinfo{author}{\bibfnamefont{I.}~\bibnamefont{Peschel}},
  \bibinfo{journal}{J. Phys. A} \textbf{\bibinfo{volume}{22}},
  \bibinfo{pages}{4257} (\bibinfo{year}{1989}).

\bibitem[{\citenamefont{Kaplan et~al.}(1990)\citenamefont{Kaplan, der Linden,
  and Horsch}}]{Kaplan1990}
\bibinfo{author}{\bibfnamefont{T.~A.} \bibnamefont{Kaplan}},
  \bibinfo{author}{\bibfnamefont{W.~V.} \bibnamefont{der Linden}},
  \bibnamefont{and} \bibinfo{author}{\bibfnamefont{P.}~\bibnamefont{Horsch}},
  \bibinfo{journal}{Phys. Rev. B} \textbf{\bibinfo{volume}{42}},
  \bibinfo{pages}{4663} (\bibinfo{year}{1990}).

\bibitem[{\citenamefont{Bernu et~al.}(1992)\citenamefont{Bernu, Lhuillier, and
  Pierre}}]{Bernu1992}
\bibinfo{author}{\bibfnamefont{B.}~\bibnamefont{Bernu}},
  \bibinfo{author}{\bibfnamefont{C.}~\bibnamefont{Lhuillier}},
  \bibnamefont{and} \bibinfo{author}{\bibfnamefont{L.}~\bibnamefont{Pierre}},
  \bibinfo{journal}{Phys. Rev. Lett.} \textbf{\bibinfo{volume}{69}},
  \bibinfo{pages}{2590} (\bibinfo{year}{1992}).

\bibitem[{\citenamefont{Bernu et~al.}(1994)\citenamefont{Bernu, Lecheminant,
  Lhuillier, and Pierre}}]{Bernu1994}
\bibinfo{author}{\bibfnamefont{B.}~\bibnamefont{Bernu}},
  \bibinfo{author}{\bibfnamefont{P.}~\bibnamefont{Lecheminant}},
  \bibinfo{author}{\bibfnamefont{C.}~\bibnamefont{Lhuillier}},
  \bibnamefont{and} \bibinfo{author}{\bibfnamefont{L.}~\bibnamefont{Pierre}},
  \bibinfo{journal}{Phys. Rev. B} \textbf{\bibinfo{volume}{50}},
  \bibinfo{pages}{10048} (\bibinfo{year}{1994}).

\bibitem[{\citenamefont{Onuki}(2002)}]{Onuki2002}
\bibinfo{author}{\bibfnamefont{A.}~\bibnamefont{Onuki}},
  \emph{\bibinfo{title}{Phase transition dynamics}}
  (\bibinfo{publisher}{Cambridge University Press}, \bibinfo{year}{2002}).

\bibitem[{\citenamefont{Sachdev}(2011)}]{Sachdev2011}
\bibinfo{author}{\bibfnamefont{S.}~\bibnamefont{Sachdev}},
  \emph{\bibinfo{title}{Quantum Phase Transitions}}
  (\bibinfo{publisher}{Cambridge University Press}, \bibinfo{year}{2011}).

\bibitem[{\citenamefont{van Wezel and van~den Brink}(2007)}]{vanWezel2007}
\bibinfo{author}{\bibfnamefont{J.}~\bibnamefont{van Wezel}} \bibnamefont{and}
  \bibinfo{author}{\bibfnamefont{J.}~\bibnamefont{van~den Brink}},
  \bibinfo{journal}{Am. J. Phys.} \textbf{\bibinfo{volume}{75}},
  \bibinfo{pages}{635} (\bibinfo{year}{2007}).

\bibitem[{\citenamefont{Birol et~al.}(2007)\citenamefont{Birol, Dereli,
  M\"{u}stecaplio\u{g}lu, and You}}]{Birol2007}
\bibinfo{author}{\bibfnamefont{T.}~\bibnamefont{Birol}},
  \bibinfo{author}{\bibfnamefont{T.}~\bibnamefont{Dereli}},
  \bibinfo{author}{\bibfnamefont{O.~E.} \bibnamefont{M\"{u}stecaplio\u{g}lu}},
  \bibnamefont{and} \bibinfo{author}{\bibfnamefont{L.}~\bibnamefont{You}},
  \bibinfo{journal}{Phys. Rev. A} \textbf{\bibinfo{volume}{76}},
  \bibinfo{pages}{043616} (\bibinfo{year}{2007}).

\bibitem[{\citenamefont{van Wezel and van~den Brink}(2008)}]{vanWezel2008sc}
\bibinfo{author}{\bibfnamefont{J.}~\bibnamefont{van Wezel}} \bibnamefont{and}
  \bibinfo{author}{\bibfnamefont{J.}~\bibnamefont{van~den Brink}},
  \bibinfo{journal}{Phys. Rev. B} \textbf{\bibinfo{volume}{77}},
  \bibinfo{pages}{064523} (\bibinfo{year}{2008}).

\bibitem[{\citenamefont{Messiah}(1999)}]{Messiah1999}
\bibinfo{author}{\bibfnamefont{A.}~\bibnamefont{Messiah}},
  \emph{\bibinfo{title}{Quantum mechanics}} (\bibinfo{publisher}{Dover
  publications}, \bibinfo{year}{1999}), ISBN \bibinfo{isbn}{9780486409245}.

\bibitem[{\citenamefont{Nielsen and Chuang}(2000)}]{Chuang2000}
\bibinfo{author}{\bibfnamefont{M.~A.} \bibnamefont{Nielsen}} \bibnamefont{and}
  \bibinfo{author}{\bibfnamefont{I.~L.} \bibnamefont{Chuang}},
  \emph{\bibinfo{title}{Quantum Computation and Quantum Information}}
  (\bibinfo{publisher}{Cambridge Uniersity Press}, \bibinfo{year}{2000}), ISBN
  \bibinfo{isbn}{9780521635035}.

\bibitem[{\citenamefont{Caldeira and Leggett}(1983)}]{Leggett1983}
\bibinfo{author}{\bibfnamefont{A.~O.} \bibnamefont{Caldeira}} \bibnamefont{and}
  \bibinfo{author}{\bibfnamefont{A.~J.} \bibnamefont{Leggett}},
  \bibinfo{journal}{Physica A} \textbf{\bibinfo{volume}{121}},
  \bibinfo{pages}{587} (\bibinfo{year}{1983}).

\bibitem[{\citenamefont{Leggett et~al.}(1987)\citenamefont{Leggett,
  Chakravarty, Dorsey, Fisher, Garg, and Zwerger}}]{Leggett1987}
\bibinfo{author}{\bibfnamefont{A.~J.} \bibnamefont{Leggett}},
  \bibinfo{author}{\bibfnamefont{S.}~\bibnamefont{Chakravarty}},
  \bibinfo{author}{\bibfnamefont{A.~T.} \bibnamefont{Dorsey}},
  \bibinfo{author}{\bibfnamefont{M.~P.~A.} \bibnamefont{Fisher}},
  \bibinfo{author}{\bibfnamefont{A.}~\bibnamefont{Garg}}, \bibnamefont{and}
  \bibinfo{author}{\bibfnamefont{W.}~\bibnamefont{Zwerger}},
  \bibinfo{journal}{Rev. Mod. Phys.} \textbf{\bibinfo{volume}{59}},
  \bibinfo{pages}{1} (\bibinfo{year}{1987}).

\bibitem[{\citenamefont{Adler}(2003)}]{Adler2003}
\bibinfo{author}{\bibfnamefont{S.~L.} \bibnamefont{Adler}},
  \bibinfo{journal}{Stud. Hist. Philos. Mod. Phys.}
  \textbf{\bibinfo{volume}{34}}, \bibinfo{pages}{135} (\bibinfo{year}{2003}).

\bibitem[{\citenamefont{Schr\"odinger}(1935)}]{Schrodinger1935}
\bibinfo{author}{\bibfnamefont{E.}~\bibnamefont{Schr\"odinger}},
  \bibinfo{journal}{Naturwissenschaften} \textbf{\bibinfo{volume}{23}},
  \bibinfo{pages}{807} (\bibinfo{year}{1935}).

\bibitem[{\citenamefont{Zurek}(1981)}]{Zurek81}
\bibinfo{author}{\bibfnamefont{W.~H.} \bibnamefont{Zurek}},
  \bibinfo{journal}{Phys. Rev. D} \textbf{\bibinfo{volume}{24}},
  \bibinfo{pages}{1516} (\bibinfo{year}{1981}).

\bibitem[{\citenamefont{van Wezel}(2010)}]{vanWezel2010}
\bibinfo{author}{\bibfnamefont{J.}~\bibnamefont{van Wezel}},
  \bibinfo{journal}{Symmetry} \textbf{\bibinfo{volume}{2}},
  \bibinfo{pages}{582} (\bibinfo{year}{2010}).

\bibitem[{\citenamefont{Mertens et~al.}(2021)\citenamefont{Mertens, Wesseling,
  Vercauteren, Corrales-Salazar, and van Wezel}}]{Mertens2021}
\bibinfo{author}{\bibfnamefont{L.}~\bibnamefont{Mertens}},
  \bibinfo{author}{\bibfnamefont{M.}~\bibnamefont{Wesseling}},
  \bibinfo{author}{\bibfnamefont{N.}~\bibnamefont{Vercauteren}},
  \bibinfo{author}{\bibfnamefont{A.}~\bibnamefont{Corrales-Salazar}},
  \bibnamefont{and} \bibinfo{author}{\bibfnamefont{J.}~\bibnamefont{van
  Wezel}}, \bibinfo{journal}{Phys. Rev. A} \textbf{\bibinfo{volume}{104}},
  \bibinfo{pages}{052224} (\bibinfo{year}{2021}).

\bibitem[{\citenamefont{Berry}(1998)}]{Berry1998}
\bibinfo{author}{\bibfnamefont{M.~V.} \bibnamefont{Berry}},
  \bibinfo{journal}{J. Phys. A: Math. Gen.} \textbf{\bibinfo{volume}{31}},
  \bibinfo{pages}{3493} (\bibinfo{year}{1998}).

\bibitem[{\citenamefont{Hatano and Nelson}(1997)}]{Hatano1997}
\bibinfo{author}{\bibfnamefont{N.}~\bibnamefont{Hatano}} \bibnamefont{and}
  \bibinfo{author}{\bibfnamefont{D.~R.} \bibnamefont{Nelson}},
  \bibinfo{journal}{Phys. Rev. B} \textbf{\bibinfo{volume}{56}},
  \bibinfo{pages}{8651} (\bibinfo{year}{1997}).

\bibitem[{\citenamefont{Yao and Wang}(2018)}]{Wang2018}
\bibinfo{author}{\bibfnamefont{S.}~\bibnamefont{Yao}} \bibnamefont{and}
  \bibinfo{author}{\bibfnamefont{Z.}~\bibnamefont{Wang}},
  \bibinfo{journal}{Phys. Rev. Lett.} \textbf{\bibinfo{volume}{121}},
  \bibinfo{pages}{086803} (\bibinfo{year}{2018}).

\bibitem[{\citenamefont{Martinez~Alvarez
  et~al.}(2018)\citenamefont{Martinez~Alvarez, Barrios~Vargas, and
  Foa~Torres}}]{Alvarez2018}
\bibinfo{author}{\bibfnamefont{V.~M.} \bibnamefont{Martinez~Alvarez}},
  \bibinfo{author}{\bibfnamefont{J.~E.} \bibnamefont{Barrios~Vargas}},
  \bibnamefont{and} \bibinfo{author}{\bibfnamefont{L.~E.~F.}
  \bibnamefont{Foa~Torres}}, \bibinfo{journal}{Phys. Rev. B}
  \textbf{\bibinfo{volume}{97}}, \bibinfo{pages}{121401}
  (\bibinfo{year}{2018}).

\bibitem[{\citenamefont{Lee and Thomale}(2019)}]{Lee2019}
\bibinfo{author}{\bibfnamefont{C.~H.} \bibnamefont{Lee}} \bibnamefont{and}
  \bibinfo{author}{\bibfnamefont{R.}~\bibnamefont{Thomale}},
  \bibinfo{journal}{Phys. Rev. B} \textbf{\bibinfo{volume}{99}},
  \bibinfo{pages}{201103} (\bibinfo{year}{2019}).

\bibitem[{\citenamefont{Bergholtz et~al.}(2021)\citenamefont{Bergholtz, Budich,
  and Kunst}}]{Bergholtz2020}
\bibinfo{author}{\bibfnamefont{E.~J.} \bibnamefont{Bergholtz}},
  \bibinfo{author}{\bibfnamefont{J.~C.} \bibnamefont{Budich}},
  \bibnamefont{and} \bibinfo{author}{\bibfnamefont{F.~K.} \bibnamefont{Kunst}},
  \bibinfo{journal}{Rev. Mod. Phys.} \textbf{\bibinfo{volume}{93}},
  \bibinfo{pages}{015005} (\bibinfo{year}{2021}).

\bibitem[{\citenamefont{Ghatak et~al.}(2020)\citenamefont{Ghatak,
  Brandenbourger, van Wezel, and Coulais}}]{Ghatak2020}
\bibinfo{author}{\bibfnamefont{A.}~\bibnamefont{Ghatak}},
  \bibinfo{author}{\bibfnamefont{M.}~\bibnamefont{Brandenbourger}},
  \bibinfo{author}{\bibfnamefont{J.}~\bibnamefont{van Wezel}},
  \bibnamefont{and} \bibinfo{author}{\bibfnamefont{C.}~\bibnamefont{Coulais}},
  \bibinfo{journal}{Proc. Nat. Acad. S.} \textbf{\bibinfo{volume}{117}},
  \bibinfo{pages}{29561} (\bibinfo{year}{2020}).

\bibitem[{\citenamefont{Bender and Boettcher}(1998)}]{Bender1998}
\bibinfo{author}{\bibfnamefont{C.~M.} \bibnamefont{Bender}} \bibnamefont{and}
  \bibinfo{author}{\bibfnamefont{S.}~\bibnamefont{Boettcher}},
  \bibinfo{journal}{Phys. Rev. Lett.} \textbf{\bibinfo{volume}{80}},
  \bibinfo{pages}{5243} (\bibinfo{year}{1998}).

\bibitem[{\citenamefont{Mostafazadeh}(2002)}]{Mostafazadeh2002}
\bibinfo{author}{\bibfnamefont{A.}~\bibnamefont{Mostafazadeh}},
  \bibinfo{journal}{J. Math. Phys.} \textbf{\bibinfo{volume}{43}},
  \bibinfo{pages}{205} (\bibinfo{year}{2002}).

\bibitem[{\citenamefont{Rodr\'{i}guez-Lara and Guerrero}(2015)}]{Lara2015}
\bibinfo{author}{\bibfnamefont{B.~M.} \bibnamefont{Rodr\'{i}guez-Lara}}
  \bibnamefont{and} \bibinfo{author}{\bibfnamefont{J.}~\bibnamefont{Guerrero}},
  \bibinfo{journal}{Opt. Lett.} \textbf{\bibinfo{volume}{40}},
  \bibinfo{pages}{5682} (\bibinfo{year}{2015}).

\bibitem[{\citenamefont{Sudarshan}(1961)}]{Sudarshan1961}
\bibinfo{author}{\bibfnamefont{E.~C.~G.} \bibnamefont{Sudarshan}},
  \bibinfo{journal}{Phys. Rev.} \textbf{\bibinfo{volume}{123}},
  \bibinfo{pages}{2183} (\bibinfo{year}{1961}).

\bibitem[{\citenamefont{Scholtz et~al.}(1992)\citenamefont{Scholtz, Geyer, and
  Hahne}}]{Scholz1992}
\bibinfo{author}{\bibfnamefont{F.~G.} \bibnamefont{Scholtz}},
  \bibinfo{author}{\bibfnamefont{H.~B.} \bibnamefont{Geyer}}, \bibnamefont{and}
  \bibinfo{author}{\bibfnamefont{F.~J.~W.} \bibnamefont{Hahne}},
  \bibinfo{journal}{Ann. Phys., NY} \textbf{\bibinfo{volume}{213}},
  \bibinfo{pages}{74} (\bibinfo{year}{1992}).

\bibitem[{\citenamefont{Penrose}(1996)}]{Penrose}
\bibinfo{author}{\bibfnamefont{R.}~\bibnamefont{Penrose}},
  \bibinfo{journal}{Gen. Relativ. Gravit.} \textbf{\bibinfo{volume}{28}},
  \bibinfo{pages}{581} (\bibinfo{year}{1996}).

\end{thebibliography}
\end{document}